\documentclass[fleqn,10pt]{wlscirep}
\usepackage[utf8]{inputenc}
\usepackage[T1]{fontenc}

\usepackage{amsmath,amssymb}
\usepackage{graphicx}
\usepackage{caption}
\usepackage{subcaption}
\usepackage{booktabs}
\usepackage{multirow}
\usepackage{url}
\usepackage{longtable}
\usepackage[table]{xcolor}
\usepackage{changes}
\makeatletter
\AddToHook{cmd/added/before}{\def\Changes@AuthorColor{red}}
\AddToHook{cmd/deleted/before}{\def\Changes@AuthorColor{red}}
\AddToHook{cmd/replaced/before}{\def\Changes@AuthorColor{red}}
\makeatother

\title{A multi-source dataset for solar energetic particle forecasting}

\author[1]{Yian Yu}
\author[1,*]{Yang Chen}
\author[2]{Lulu Zhao}
\author[3]{Kathryn Whitman}
\author[2]{Ward Manchester}
\author[2]{Tamas Gombosi}

\affil[1]{Department of Statistics, University of Michigan, Ann Arbor, MI 48109, USA}
\affil[2]{Department of Climate and Space Sciences and Engineering, University of Michigan, Ann Arbor, MI 48109, USA}
\affil[3]{NASA Space Radiation Analysis Group, Johnson Space Center, Houston, TX 77058, USA}

\affil[*]{Corresponding author: ychenang@umich.edu}


\begin{abstract}
Solar energetic particle (SEP) event forecasting often involves integrating heterogeneous observations that differ in cadence, temporal coverage, format, and historical availability, posing challenges for reproducible analysis of data-driven approaches. This paper presents \texttt{SEP-PRISM Data}, a curated multi-source dataset designed for 24-hour-ahead forecasting of operational SEP events, defined as proton flux exceeding 10 pfu in the GOES $>$ 10 MeV channel. \texttt{SEP-PRISM Data} integrates flare records, active-region magnetic field parameters, coronal mass ejection (CME) catalog data, GOES soft X-ray flux, and historical proton flux into a common window-based representation spanning 3 February 1986 to 10 September 2025. To improve temporal coverage and cross-source consistency, SHARP and SMARP magnetic products were aligned into a unified SMHARP archive, and CME records from DONKI and CDAW were aligned into a unified CDAWDONKI event set. Predictor variables were summarized over fixed non-overlapping 24-hour historical windows using minimum, mean, and maximum statistics and paired with targets defined over the subsequent 24-hour window, forming a supervised learning dataset. The resulting \texttt{SEP-PRISM Data} contains 14,464 labeled samples, including 650 positive operational SEP cases, and is intended to support reproducible benchmarking, model development, feature analysis, and future studies of space weather forecasting. 
\end{abstract}
\begin{document}


\flushbottom
\maketitle
%
%
\thispagestyle{empty}


\section*{Background \& Summary}
Solar energetic particle (SEP) events are among the most important space weather hazards because they can endanger astronauts, damage spacecraft systems, and disrupt high-latitude aviation and technological infrastructure \cite{Eastwood2017Economic,whitman2024multi}. Despite decades of studies, SEP forecasting remains difficult. SEP events are rare, depend on multiple solar and heliospheric drivers, including flare- and coronal mass ejection (CME)-associated acceleration, magnetic connectivity, and interplanetary propagation, and vary substantially in onset, intensity, and duration \cite{REAMES2004,KIM2011,desai2016large,klein2017acceleration}. Recent reviews further show that reported skill {scores} {vary} across studies, in part because of uneven data coverage, inconsistent event definitions, and the lack of harmonized datasets suitable for reproducible benchmarking and operational validation \cite{Whitman2023Review,sadykov2025,kasapis2026review}.

These limitations motivated the development \texttt{SEP-PRISM Data}, which is used in \texttt{SEPNET-PRISM Model}\cite{yu2026OperationalSEP}. The dataset was constructed to provide a unified resource for 24-hour ahead forecasting of operational SEP events, defined according to the National Oceanic and Atmospheric Administration (NOAA) operational criterion of proton flux exceeding 10 pfu in the Geostationary Operational Environmental Satellite (GOES) $>$ 10 MeV channel \cite{NOAA_SWC_ProtonFlux}. In addition to the primary operational target, the dataset also retains the broader general SEP events, which corresponds to a more inclusive SEP threshold, such as proton flux enhanced above the GOES background level (indicated by an arbitrarily low threshold of 10$^{-6}$ pfu) in the $>$ 10 MeV channel, and may be useful as an auxiliary target for multi-task learning \cite{Yu2026SEPNET}. SEP event labels were taken from the CLEAR SEP benchmark dataset \cite{fetchsep}, which provides an expanded list of SEP events compiled from February 1986 through September 2025. In that benchmark, particle and detector background levels were identified and set to zero, leaving non-zero fluxes only during enhanced SEP periods \cite{fetchsep}. To support data-driven forecasting studies, we combine {active}-region magnetic parameters, flare records, CME catalogs, GOES soft X-ray flux, and historical proton flux into a common supervised learning format.

A central motivation for \texttt{SEP-PRISM Data} is the limited temporal coverage of key observational sources for long-horizon SEP event forecasting. Magnetic predictors from Space-weather HMI Active Region Patch (SHARP) begin only in 2010 \cite{Bobra2021SMARPSHARP}, whereas flare and proton observations extend much further back in time. Likewise, Database of Notifications, Knowledge, Information (DONKI) CME records do not cover the earlier solar cycles needed for long-range benchmarking \cite{ccmc_donki}. The curation workflow, therefore, combines original observations with aligned surrogate products, producing two key extended records: SMHARP, which extends SHARP-like magnetic predictors back to 1996 using Space-weather MDI Active Region Patch (SMARP) observations \cite{Bobra2021SMARPSHARP}, and CDAWDONKI, which extends DONKI-like CME attributes back to 1996 using the Coordinated Data Analysis Workshop (CDAW) CME catalog \cite{gopalswamy2025CDAW}.

The resulting \texttt{SEP-PRISM Data} are not simply a collection of source files, but a {data product structured as paired predictor-label-samples ready for direct use in supervised machine learning}. Heterogeneous inputs are aggregated into fixed, non-overlapping 24-hour windows, and each predictor is represented by its three summary statistics: the minimum, mean, and maximum values over the historical 24-hour window. These summarized predictors are paired with labels defined over the subsequent 24 hours, including a binary operational SEP target and several auxiliary regression targets, such as flare count, CME count, maximum CME speed, proton flux, and GOES soft X-ray flux, over the same period. This design makes the database immediately usable for machine-learning benchmarks, model comparison, feature analysis, and operationally motivated SEP forecasting studies while preserving provenance from the underlying data streams. The released dataset has also been used in SEP forecasting studies for developing real-time forecasting models, with encouraging predictive performance reported in the companion work \cite{yu2026OperationalSEP}.

\section*{Methods}

\subsection*{Source data preprocessing and alignment}
We describe the different components of \texttt{SEP-PRISM Data} here, which is used to train a real-time SEP forecasting model in \texttt{SEPNET-PRISM Model}\cite{yu2026OperationalSEP}. Part of the data description is available in the \texttt{SEPNET-PRISM Model}\cite{yu2026OperationalSEP} paper, but we provide a more detailed account here of the source-data acquisition, preprocessing, alignment, and construction of the released data products.

\begin{figure}[hpbt]
    \centering
    \includegraphics[width=1\linewidth]{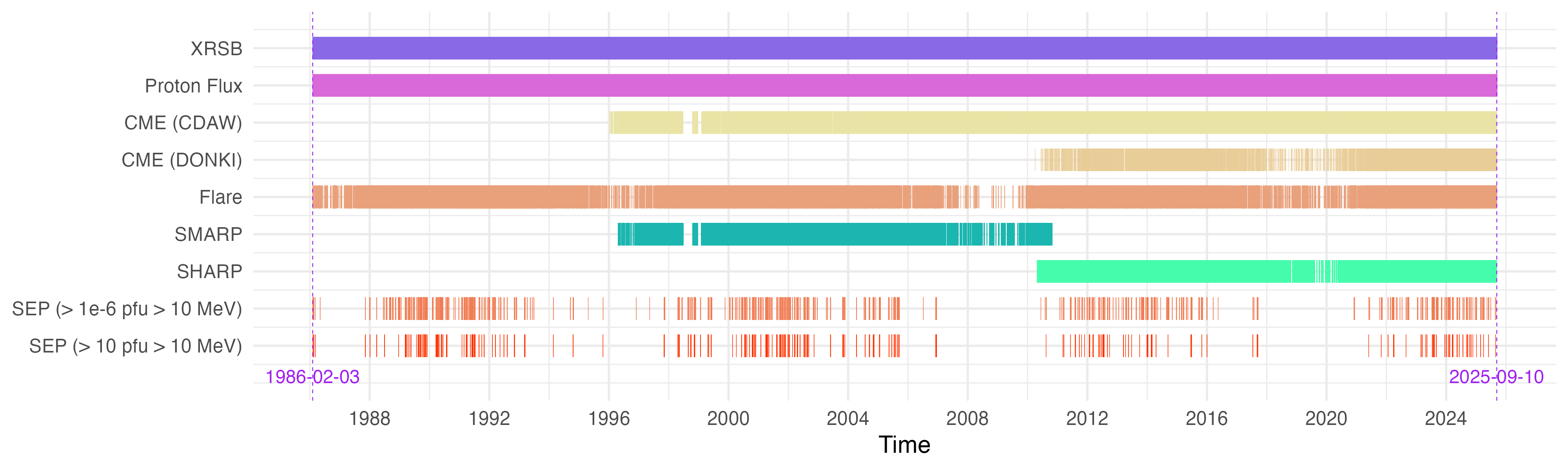}
    \caption{Timeline of multi-source data used in this study from 3 February 1986 to 10 September 2025 (adapted from a figure originally published in the Supplementary Information of \texttt{SEPNET-PRISM Model} \cite{yu2026OperationalSEP}). The rows show GOES X-ray flux (XRSB), proton flux ($>$ 10 MeV), flares,  CDAW/DONKI CME events, SHARP/SMARP parameters, and SEP events. Shaded bars indicate data availability and the start time and end times of events (flares, CMEs, and SEPs). The bottom two rows show general SEP events (proton flux $>$ $1\times 10^{-6}$ pfu in the $>$ 10 MeV channel) and operational SEP events (proton flux $>$ 10 pfu in the $>$ 10 MeV channel), respectively. }
    \label{fig:multi_source_timeline}
\end{figure}

The primary SEP event records were obtained from the CLEAR SEP benchmark dataset, which spans 3 February 1986 to 10 September 2025 and contains 569 entries \cite{fetchsep} {The catalog was downloaded from the public \texttt{fetchsep} GitHub release at \url{{https://github.com/ktindiana/fetchsep/releases/tag/CLEAR_Benchmark_v1.0}}}. Of the 569 entries, 267 correspond to operational SEP events, defined as proton flux exceeding 10 pfu in the $>$ 10 MeV GOES integral primary channel; 567 entries exceed the more general low threshold of $1\times 10^{-6}$ pfu. {The remaining 2 entries have a recorded maximum flux but lack the corresponding start/end times, duration, onset peak, and fluence fields, and are therefore excluded from event-level statistics}. {The general threshold of 10$^{-6}$ pfu follows the background subtraction procedure used in the CLEAR SEP Benchmark pipeline \cite{fetchsep}. Instrumental and particle background levels are estimated and set to zero; any nonzero residual flux above this near-floor value is treated as an enhancement. This threshold also provides additional positive labels for the imbalanced operational SEP task; the previous \texttt{SEPNET} model exploits this by training on general threshold events and then calibrating and evaluating on the operational threshold \cite{Yu2026SEPNET}.} The database supports two binary classification targets: operational SEP events ($\geq$ 10 pfu in $>$ 10 MeV channel) and general SEP events ($\geq$ $1\times10^{-6}$ pfu in $>$ 10 MeV channel) \cite{Yu2026SEPNET}. The predictor set integrates five data families: flare records, SMHARP magnetic parameters and flare-matched SMHARP subsets, CDAWDONKI CME records, GOES soft X-ray flux (XRSB), and proton flux in the GOES $>$ 10 MeV channel. The observational data sources are summarized in Table \ref{tab:source_data} and illustrated in Figure \ref{fig:multi_source_timeline}. {Except for the CLEAR SEP Benchmark dataset, which was downloaded directly from the public \texttt{fetchsep} GitHub release, all source datasets were retrieved programmatically using \texttt{Code/fetch\_solar\_data.ipynb}, archived with this dataset at Zenodo \cite{Yu2026SEPPRISMData} (\url{https://doi.org/10.5281/zenodo.21297634}). It records the source endpoints, requested data products or series, query parameters, date ranges, and retrieval dates. Downloaded external raw input records were used locally during data construction and are not redistributed in the Zenodo archive.}

\begin{table}[hpbt]
\centering
\caption{Summary of the observational data sources.}
\label{tab:source_data}
\begin{tabular}{p{3.5cm} p{4.5cm} p{4cm} p{1.8cm}p{1.5cm}}
\hline
{Data source} & {Main variables used} & {Time coverage} & {Cadence}& {Records} \\
\hline
CLEAR SEP Benchmark & Operational SEP event start and end times & 3 Feb 1986-10 Sep 2025 & Event-based & 569\\
\rowcolor{lightgray!30} Flare & Duration, rise time, log strength,  NOAA active-region number& 3 Feb 1986-10 Sep 2025 & Event-based  & 78,724 \\
XRSB & Soft X-ray flux & 3 Feb 1986-10 Sep 2025 & 5 min & 4,165,920\\
\rowcolor{lightgray!30} Proton flux ($>$ 10 MeV) & Flux value& 3 Feb 1986-10 Sep 2025 & 5 min & 4,165,920 \\
SHARP & Magnetic field parameters and active-region descriptors & 1 May 2010-10 Sep 2025 & 12 min & 4,889,762\\
\rowcolor{lightgray!30} SMARP & Magnetic field parameters and active-region descriptors & 23 Apr 1996-27 Oct 2010 & 96 min & 68,166 \\
DONKI CME catalog& Radial speed, longitude, half angular width & 3 Apr 2010-9 Sep 2025 & Event-based & 8,366\\
\rowcolor{lightgray!30} CDAW CME catalog& Radial speed, central position angle, angular width, kinetic energy & 11 Jan 1996-9 Sep 2025 & Event-based & 16,216\\
\hline
\end{tabular}
\end{table}

\paragraph{Flare records.}
Flare events were retrieved from the Heliophysics Event Knowledgebase (HEK) through SunPy Fido  \cite{Barnes2020SunPy,hek_dataset} {The HEK service is available at \url{https://www.lmsal.com/hek/}. In
\texttt{Code/fetch\_solar\_data.ipynb}, flare events were queried over the
specified UTC date interval using \texttt{a.Time(start, end)},
\texttt{a.hek.EventType("FL")}, the constraint
\texttt{a.hek.FL.GOESCls > "A1.0"}, and the observatory constraint
\texttt{a.hek.OBS.Observatory == "GOES"}}. The resulting catalog contains 78,724 records over the same period as the SEP benchmark. For each event, the retained fields include start, peak, and end times, GOES flare class, associated NOAA active-region number, and heliographic Stonyhurst coordinates. Additional descriptors were derived during preprocessing: duration (minutes from start to end), risen time (minutes from start to peak), peak flux strength in W\,m$^{-2}$ and logarithmic peak strength. {Duration and rise time were computed directly from the timestamp differences; peak flux strength was converted from the reported GOES flare class using the standard class-to-flux convention (letter class sets the decade from $10^{-8}$ to $10^{-4}$ W\,m$^{-2}$, scaled by the numeric coefficient), and logarithmic peak strength is $\log_{10}$ of this value.} These features provide a compact summary of flare timing and intensity and are later aggregated into daily-window statistics.

\paragraph{Active-region magnetic fields.} Space-based observations of solar photospheric magnetic fields from the Helioseismic and Magnetic Imager on board Solar Dynamics Observatory (SDO/HMI) \cite{scherrer2012helioseismic} and the Michelson Doppler Imager on board Solar and Heliospheric Observatory (SOHO/MDI) \cite{scherrer1995solar} provide continuous monitoring of the line-of-sight and vector magnetic fields. SHARP and SMARP are data products derived from these observations, providing summary parameters and maps for automatically tracked solar active regions \cite{Bobra2021SMARPSHARP}. Both data products were retrieved from the publicly Joint Science Operations Center (JSOC) {archive at Stanford University (\url{http://jsoc.stanford.edu/ajax/lookdata.html})} using the drms client \cite{Glogowski2019}. SHARP data were retrieved from the definitive vector-magnetic-field series \textit{hmi.sharp\_cea\_720s}, which is available at 12-minute cadence from May 2010 onward and is remapped to a cylindrical equal-area projection. When definitive SHARP products were unavailable, records were retrieved from the near-real-time series \textit{hmi.sharp\_cea\_720s\_nrt}. SMARP records were retrieved from the \textit{mdi.smarp\_cea\_96m} series, which covers 23 April 1996 to 27 October 2010 at 96-minute cadence. The exact JSOC record-set queries, requested keywords, date ranges, and retrieval procedures are provided in \texttt{Code/fetch\_solar\_data.ipynb} in the Zenodo archive.

The retrieved magnetic parameters, summarized in Table \ref{tab:sharp_parameters}, include unsigned magnetic flux, mean field components, current and helicity measures, free-energy proxies, shear statistics, size and area metrics, and heliographic location descriptors, with a subset shared between SHARP and SMARP and additional SHARP-specific parameters reconstructed through alignment models. The resulting unified magnetic archive is referred to as SMHARP. In addition to the complete archive, a flare-matched subset was created by linking magnetic records to flare-producing NOAA active-region numbers whenever a shared active-region identifier was available or a spatial proximity criterion supported the association. This matched subset provides localized magnetic information tied to flare-producing regions, while the full SMHARP archive retains broader contextual information on active-region magnetic conditions over time.

\begin{longtable}{lcc}
\caption{Description of the SHARP and SMARP parameters. Parameters marked with $^\ast$ are available in both SHARP and SMARP datasets. The active-region identifiers \texttt{HARPNUM} in SHARP and \texttt{TARPNUM} in SMARP were unified and denoted as \texttt{ARPNUM}.}
\label{tab:sharp_parameters} \\
\hline
Feature & Description & Unit \\ \hline

\texttt{USFLUX}   & Total unsigned magnetic flux & Mx \\
\rowcolor{lightgray!30} \texttt{MEANGAM}  & Mean inclination angle & degree \\
\texttt{MEANGBT}  & Mean total field gradient & G\,Mm$^{-1}$ \\
\rowcolor{lightgray!30} \texttt{MEANGBZ}  & Mean vertical field gradient & G\,Mm$^{-1}$ \\
\texttt{MEANGBH}  & Mean horizontal field gradient & G\,Mm$^{-1}$ \\
\rowcolor{lightgray!30} \texttt{MEANJZD}  & Mean vertical current density & mA\,m$^{-2}$ \\
\texttt{TOTUSJZ} & Total unsigned vertical current & A \\
\rowcolor{lightgray!30} \texttt{MEANALP} & Mean twist parameter $\alpha$ & Mm$^{-1}$ \\
\texttt{MEANJZH}  & Mean current helicity & G$^{2}$\,m$^{-1}$ \\
\rowcolor{lightgray!30}\texttt{SAVNCPP}  & Sum of absolute net current per polarity & A \\
\texttt{MEANPOT}  & Mean photospheric magnetic energy density & erg\,cm$^{-3}$ \\
\rowcolor{lightgray!30}\texttt{TOTPOT}   & Total photospheric magnetic energy density & erg\,cm$^{-1}$ \\
\texttt{MEANSHR}  & Mean shear angle & degree \\
\rowcolor{lightgray!30}\texttt{SHRGT45}  & Fraction of pixels with shear angle $>45^\circ$ & -- \\
\texttt{USFLUXL$^\ast$} & Line-of-sight unsigned flux & Mx \\
\rowcolor{lightgray!30}\texttt{MEANGBL$^\ast$} & Mean gradient of line-of-sight field & G\,Mm$^{-1}$ \\
\texttt{ARPNUM$^\ast$}  & Active-region patch number & -- \\
\rowcolor{lightgray!30}\texttt{NPIX$^\ast$}    & Total number of pixels & count \\
\texttt{SIZE$^\ast$}   & Total patch size & pixel \\
\rowcolor{lightgray!30}\texttt{AREA$^\ast$}    & Total area of patch & Mm$^{2}$ \\
\texttt{NACR$^\ast$}    & Number of active pixels & count \\
\rowcolor{lightgray!30}\texttt{SIZE\_ACR$^\ast$} & Size of active-region pixels & pixel \\
\texttt{AREA\_ACR$^\ast$} & Area of active-region pixels & Mm$^{2}$ \\
\rowcolor{lightgray!30}\texttt{LAT\_MIN$^\ast$} & Minimum latitude of bounding box & degree \\
\texttt{LON\_MIN$^\ast$} & Minimum longitude of bounding box & degree \\
\rowcolor{lightgray!30}\texttt{LAT\_MAX$^\ast$} & Maximum latitude of bounding box & degree \\
\texttt{LON\_MAX$^\ast$} & Maximum longitude of bounding box & degree \\
\rowcolor{lightgray!30}\texttt{LAT\_FWT$^\ast$} & Flux-weighted latitude & degree \\
\texttt{LON\_FWT$^\ast$} & Flux-weighted longitude & degree \\
\rowcolor{lightgray!30}\texttt{CRLT\_OBS$^\ast$} & Carrington latitude of observer & degree \\
\texttt{CRLN\_OBS$^\ast$} & Carrington longitude of observer & degree \\

\rowcolor{lightgray!30}\texttt{R\_VALUE$^\ast$}  & Sum of flux near polarity inversion lines & Mx \\
\texttt{NOAA\_AR$^\ast$}  & NOAA active region number & -- \\
\rowcolor{lightgray!30}\texttt{CAR\_ROT$^\ast$}  & Carrington rotation number & -- \\
\texttt{CMASKL$^\ast$}    & Bitmap mask for active-region pixels & -- \\
\rowcolor{lightgray!30}\texttt{DSUN\_OBS$^\ast$} & Sun-observer distance & m \\
\texttt{RSUN\_OBS$^\ast$} & Apparent solar radius & arcsec \\
\hline
\end{longtable}

\paragraph{CME records.}
CME predictors were assembled from the NASA DONKI \cite{ccmc_donki} and the
SOHO/LASCO CME Catalog maintained by the CDAW Data Center \cite{CDAW_LASCO_CME}. {DONKI CME records were retrieved from the public endpoint \url{https://kauai.ccmc.gsfc.nasa.gov/DONKI/WS/get/CME}. CDAW records, derived from SOHO/LASCO coronagraph observations, were retrieved from the Version 2 catalog endpoint
\url{https://cdaw.gsfc.nasa.gov/CME_list/UNIVERSAL_ver2}.} DONKI provides event records from 3 April 2010 onward with activity identifier, start time, source location, half angular width, and radial speed. CDAW extends coverage back to 1996 with event time, central position angle, angular width, linear speed, and kinetic energy. After removal of low-quality CDAW entries flagged as \emph{Poor} or \emph{Very Poor}, the two catalogs were aligned so that reconstructed CDAW-based features could extend DONKI-like CME information to earlier years. The unified CME archive is referred to as CDAWDONKI. The retrieved CME parameters are summarized in Table \ref{tab:CME_features}. {The alignment procedure is described later, and its performance is evaluated in Table \ref{tab:CME_alignment_performance}.}

\begin{longtable}{lcc}
\caption{Description of the CME features. The shared feature between CDAW and DONKI is marked with $^\ast$.}
\label{tab:CME_features} \\
\hline
Feature & Description & Unit \\ \hline
\texttt{lat\_deg} & CME latitude & degree \\
\rowcolor{lightgray!30}\texttt{lon\_deg} & CME longitude & degree \\
\texttt{halfAngle\_deg} & CME half angular width & degree \\
\rowcolor{lightgray!30}\texttt{central\_pa\_deg} & Central position angle & degree \\
\texttt{width\_deg} & Angular width & degree \\
\rowcolor{lightgray!30}\texttt{energy\_erg} & CME kinetic energy & erg \\
\texttt{speed\_km\_s$^\ast$} & CME radial speed & km\,s$^{-1}$ \\
\hline
\end{longtable}

\paragraph{GOES X-ray and proton flux.}
GOES soft X-ray sensor data in the 1-8 $\mathring{A}$ channel (XRSB) and integral proton flux in the $>$ 10 MeV channel were assembled as continuous time series with 5-minute cadence. {Historical XRS observations were retrieved across
available GOES satellites using SunPy Fido with the time-range constraint
\texttt{a.Time(start, end)} and the X-ray Sensor constraint \texttt{a.Instrument("XRS")}. The retained XRSB channel corresponds to the
1--8 \AA\ (0.1--0.8 nm) soft X-ray band.} Recent XRS data gaps were
supplemented, when necessary, using the NOAA Space Weather Prediction Center
(SWPC) GOES JSON service
(\url{https://services.swpc.noaa.gov/json/goes/}). {The primary/secondary instrument-to-satellite mapping was obtained from the SWPC instrument-source metadata
(\url{https://services.swpc.noaa.gov/json/goes/instrument-sources.json}).} Measurements from each satellite were aggregated to 5-minute means and
stitched onto a common timeline according to the designated primary satellite
intervals. {At each 5-minute time step, the designated primary satellite was
used; concurrent measurements from other satellites were not averaged. The
contributing satellite identifier was retained in the merged record.}

GOES integral proton flux in the $>$10~MeV channel was retrieved through the NASA Integrated Space Weather Analysis (ISWA) HAPI service (\url{https://iswa.ccmc.gsfc.nasa.gov/hapi/}), We queried the 5-minute GOES
primary and secondary solar-particle products,
\texttt{goesp\_part\_flux\_P5M} and \texttt{goess\_part\_flux\_P5M},
respectively. From each product, we retained the \texttt{P10} parameter,
representing the integral proton flux in the $>$10~MeV channel, with units of
protons\,cm$^{-2}$\,s$^{-1}$\,sr$^{-1}$ (pfu). The corresponding \texttt{satelliteProton} parameter was retained to identify the contributing GOES satellite. The primary or secondary product was selected according to
the GOES primary/secondary instrument-status information for the relevant time interval, and the selected records were merged to form a continuous 5-minute proton-flux time series. Exact HAPI dataset identifiers, parameter names, request parameters, date ranges, retrieval dates, and processing procedures
are documented in \texttt{Code/fetch\_solar\_data.ipynb}. 

\subsection*{Alignment of magnetic records}

Before alignment, missing or invalid values in SHARP and SMARP were handled at the record level. Non-physical zero values or non-finite entries in selected magnetic variables were treated as missing. Rows that were entirely missing across the core magnetic and positional descriptors were removed. Remaining missing values were imputed within each active region using the nearest available observation in time, implemented through a rolling nearest-neighbor join on \texttt{ARPNUM} and observation time. added{Specifically, each active region's records were grouped by an exact match on \texttt{ARPNUM}, and remaining missing values were filled from the temporally nearest observation in the same active region; values were not interpolated across active regions.}

The overlapping SHARP-SMARP period in 2010 was partitioned into 96-minute windows to match the coarser SMARP cadence. Within each window, SMARP records were matched to SHARP records first by NOAA active-region identifier \texttt{NOAA\_AR} when available and unambiguous. When a direct match was not possible, records were linked by spatial proximity using either flux-weighted centers (\texttt{LAT\_FWT}, \texttt{LON\_FWT}) or bounding-box centers, defined as $\big((\texttt{LAT\_MIN}+\texttt{LAT\_MAX})/2, (\texttt{LON\_MIN}+\texttt{LON\_MAX})/2\big)$ with a five-degree angular threshold. When multiple SHARP candidates were linked to a single SMARP record, their values were averaged within the corresponding 96-minute window before model fitting.

Alignment was then performed feature by feature using regression models that map SMARP predictors to SHARP parameters. Shared SHARP parameters were reconstructed first, and SHARP-specific parameters were then modeled using both the original SMARP covariates and the fitted shared SHARP estimates. For each response feature, a full linear model was considered and the final specification was selected by stepwise AIC (Akaike Information Criterion, an information criterion for model selection\cite{akaike1974new}), choosing the model with the lowest AIC among the candidates. The fitted models were applied to pre-2010 SMARP observations to produce SHARP-equivalent values, which were then combined with native SHARP observations to form SMHARP dataset, spanning 1996 to the end of study period with 5,310,881 records.

\subsection*{Alignment of CME records}\label{sec:cme_alignment}

The overlap period between DONKI and CDAW spans 3 April 2010 to 9 September 2025. CDAW was treated as the reference catalog, and for each CDAW CME event, the nearest DONKI event was identified by absolute time difference; CDAW events were considered unmatched and excluded from the alignment set if the closest DONKI event was separated by more than 24 hours. {The bound was set to match the 24-hour window granularity used elsewhere in the dataset's supervised-learning representation, so that accepted matches are no more separated in time than the aggregation window used for forecasting.} Prior to fitting a model to the data, missing and nonphysical values, such as zero-valued entries in variables like speed and energy, were imputed within each catalog using a k-nearest neighbor method with the default setting of k = 5. Unlike the continuously sampled SHARP and SMARP parameters, CME catalogs consist of discrete events; therefore, imputation was performed in a multivariate feature space by identifying nearest-neighbor events with similar characteristics, rather than through continuous temporal (linear) interpolation.

The alignment again used regression-based reconstruction. The shared speed variable was modeled first, and DONKI-specific variables, such as longitude and half-angle, were reconstructed using CDAW features together with the fitted shared feature. Latitude showed poor agreement after alignment and was excluded from the final downstream CME feature set. The resulting unified archive, CDAWDONKI, extends DONKI-like CME information back to 1996 and contains 17,463 records.

\subsection*{Summary statistics window and target construction}
{The released analyzed datasets were constructed from the raw and processed source records through the workflow summarized in Table \ref{tab:processing-pipeline}. First, \texttt{fetch\_solar\_data.ipynb} retrieves the publicly available source records and stores them in \texttt{Raw Data/}. Source-specific preprocessing scripts then generate harmonized tables of processed data for active-region magnetic fields, flare records, CME records, GOES soft X-ray flux, and GOES $>$10~MeV proton flux. The script \texttt{All\_features\_RData.R} combines these processed tables into \texttt{All\_features.RData}, which is used together with the CLEAR SEP Benchmark catalog by \texttt{Data\_Aggregation.R} to construct the final analyzed rolling-window datasets. The plotting scripts listed in Table \ref{tab:processing-pipeline} were used only to generate the corresponding figures.}

\begin{table}[htbp]
	\centering
	\caption{Data-processing workflow used to construct the released files. All scripts are located in \texttt{Code/} and are executed from the repository root after setting \texttt{REPO\_ROOT}. Steps 0--3 generate the released data products; Steps 4--5 generate figures only.}
	\label{tab:processing-pipeline}
	\begin{tabular}{@{}cp{6.5cm}p{9cm}@{}}
		\toprule
		Step & Script  & Output(s) \\
		\midrule
		0 & \texttt{Code/fetch\_solar\_data.ipynb}  & \texttt{Raw Data/*\_raw.csv}; \mbox{\texttt{Raw Data/GOES-ProtonFlux/}}; \mbox{\texttt{Raw Data/GOES-XRS/}} \\
		 \rowcolor{lightgray!30} 1a & \texttt{Code/Dataset\_Preprocessing/} \texttt{sharp-smarp\_fusion.R}  & \texttt{Processed Data/SMHARP.csv}; \mbox{\texttt{Code/Plot/SHARP\_SMARP\_Plot.RData}} \\
		1b & \texttt{Code/Dataset\_Preprocessing/} \texttt{Flare-feature extraction.R}  & \texttt{Processed Data/Flare.csv} \\
		\rowcolor{lightgray!30} 1c & \texttt{Code/Dataset\_Preprocessing/} \texttt{CDAW\_DONKI-fusion.R} & \texttt{Processed Data/CDAWDONKI\_CME.csv}; \mbox{\texttt{Processed Data/CDAW\_CME.csv}} \\
		1d & \texttt{Code/Dataset\_Preprocessing/} \texttt{GOES-Sats\_Fusion.R}  & \texttt{Processed Data/Merged\_XRays.csv}; \mbox{\texttt{Processed Data/Merged\_ProtonFlux10.csv}} \\
		\rowcolor{lightgray!30} 1e & \texttt{Code/Dataset\_Preprocessing/} \texttt{GOES-HAPI-proton-flux\_fusion.R}  & \texttt{Processed Data/GOESHAPI\_ProtonFlux.csv} \\
		2 & \texttt{Code/All\_features\_RData.R}  & \texttt{Code/All\_features.RData} \\
		\rowcolor{lightgray!30} 3 & \texttt{Code/Data\_Aggregation.R} & \texttt{Analyzed Data/rolling\_combinded\_seq\_*.csv} \\
        4 & \texttt{Plot/dataspan.R}  & \texttt{Code/Plot/Figures/Raw\_Data.png} \\
\rowcolor{lightgray!30}
5 & \texttt{Code/Plot/sharp\_smarp\_plot.R} & \texttt{Code/Plot/Figures/SHARP\_SMARP.png} \\
		\bottomrule
	\end{tabular}
\end{table}

All heterogeneous observations were transformed into a common supervised-learning representation based on a 24-hour historical predictor window followed by a 24-hour prediction window. For an analyzed sample indexed by time \(t\), predictor variables were calculated from observations in the historical interval \([t, t+24~\mathrm{h})\), and outcome variables were calculated from the subsequent interval \([t+24~\mathrm{h}, t+48~\mathrm{h})\). For each numeric predictor available in the historical window, \texttt{Data\_Aggregation.R} calculated the minimum, arithmetic mean, and maximum. Event-based sources additionally contributed event-count variables and binary occurrence indicators. These summaries were combined into one predictor vector for each historical window.

The primary binary target, \texttt{Future\_OSEP\_label}, was assigned a value of 1 when the subsequent 24-hour prediction window overlapped at least one operational SEP event in the CLEAR SEP Benchmark catalog, defined by proton flux exceeding 10~pfu in the GOES $>$10~MeV channel. The auxiliary binary target, \texttt{Future\_GSEP\_label}, was defined analogously using the general SEP threshold of \(1\times10^{-6}\)~pfu. Additional future-window targets included the number of flare and CME events, maximum CME speed, maximum GOES $>$10~MeV proton flux, and maximum GOES 0.1--0.8~nm soft X-ray flux.

Two analyzed data products were created from this procedure. In \texttt{rolling\_combinded\_seq\_24hours.csv}, the start time \(t\) advances in 24-hour increments; thus, consecutive historical predictor windows do not overlap. In \texttt{rolling\_\\combinded\_seq\_1hours.csv}, the start time \(t\) advances in 1-hour increments, yielding overlapping 24-hour predictor windows for higher-frequency forecasting analyses. Both files use the same feature definitions, target definitions, aggregation rules, and column-naming conventions.

{Missingness was handled separately for each source type during aggregation. If no SHARP or SMHARP record was available within a historical window, \texttt{SHARP\_label} was set to 0, and all SHARP-derived summary variables for that window were recorded as \texttt{NA}. If no flare or CME event occurred within a historical window, the corresponding event-count variable was set to 0 and the associated binary occurrence indicator was set to 0. For flare windows with no event, the flare log-strength summary used the default value \(\log_{10}(10^{-10})\), as implemented in \texttt{Data\_Aggregation.R}. For continuous-flux variables, the available measurements within each window were summarized using the same minimum, mean, and maximum procedure. The complete implementation of these aggregation, labeling, and missing-data rules is available in \texttt{Code/Data\_Aggregation.R}.}

\section*{Data Records}
{The SEP-PRISM Data dataset is available at Zenodo \cite{Yu2026SEPPRISMData} (\url{https://doi.org/10.5281/zenodo.21297634}) and released under the Creative Commons Attribution 4.0 International License.} The archive contains {curated, processed, and analyzed data products}, an R workspace file, Python and R processing scripts, plotting scripts, {and source-data retrieval documentation}. All timestamps in the released tabular files are in coordinated universal time (UTC), and the database spans from 1986-02-03 00:00:00 UTC to 2025-09-10 00:00:00 UTC.

The top-level archive is organized into four directories: \texttt{Raw Data/}, \texttt{Processed Data/}, \texttt{Analyzed Data/}, and \texttt{Code/}. Table \ref{tab:data-tiers} summarizes their contents. The \texttt{Raw Data/} directory is retained as a local working directory for external source data. When
\texttt{Code/fetch\_solar\_data.ipynb} is executed, it retrieves source records
from the original providers and writes them locally to \texttt{Raw Data/},
including the CLEAR SEP Benchmark catalog, HEK flare records, JSOC SHARP and
SMARP records, DONKI and CDAW CME catalogs, and GOES X-ray and proton-flux
files. The file names and expected local outputs are documented in the
retrieval notebook and the \texttt{Raw Data/README.md} file.

\begin{table}[htbp]
  \centering
  \caption{Directory structure of the Zenodo archive.}
  \label{tab:data-tiers}
  \begin{tabular}{@{}p{2cm}p{2.5cm}p{12cm}@{}}
    \toprule
    Directory & Format & Contents \\
    \midrule
    \texttt{Raw Data/} & Markdown; local CSV files after retrieval &
README documentation and a local working directory for external source records. The downloaded raw files are generated locally by
\texttt{Code/fetch\_solar\_data.ipynb}. \\
    \rowcolor{lightgray!30}
    \texttt{Processed Data/} & CSV, RData & Harmonized source tables, merged GOES time series, and the \texttt{All\_features.RData} workspace. \\
    \texttt{Analyzed Data/} & CSV & Model-ready rolling-window feature tables and future-window target variables. \\
    \rowcolor{lightgray!30}
    \texttt{Code/} & IPYNB, R & Data-retrieval, preprocessing, aggregation, validation, and figure-generation scripts. \\
    \bottomrule
  \end{tabular}
\end{table}

\paragraph{Processed data.}
The \texttt{Processed Data/} directory contains the files listed in Table~\ref{tab:processed-inventory}.

\begin{table}[htbp]
  \centering
  \caption{Files in \texttt{Processed Data/}.}
  \label{tab:processed-inventory}
  \small
  \begin{tabular}{@{}p{5.8cm}p{10.9cm}@{}}
    \toprule
    File & Contents \\
    \midrule
    \texttt{SMHARP.csv}
      & Unified SHARP/SMARP active-region magnetic-field table. \\
    \rowcolor{lightgray!30}
    \texttt{Flare.csv}
      & Flare-event table with event timing and flare-strength variables. \\
    \texttt{CDAWDONKI\_CME.csv}
      & Unified DONKI/CDAW CME event table. \\
    \rowcolor{lightgray!30}
    \texttt{CDAW\_CME.csv}
      & CDAW CME event table with CDAW-specific attributes. \\
    \texttt{GOESHAPI\_ProtonFlux.csv}
      & Merged GOES $>$10~MeV proton-flux time series. \\
    \rowcolor{lightgray!30}
    \texttt{Merged\_XRays.csv}
      & Merged GOES 0.1--0.8~nm soft X-ray flux time series. \\
    \texttt{All\_features.RData}
      & R workspace containing the processed data tables and feature objects used by the aggregation scripts. \\
    \bottomrule
  \end{tabular}
\end{table}

In \texttt{SMHARP.csv}, \texttt{T\_REC\_posix} denotes observation time; \texttt{ARPNUM} is the unified active-region patch identifier; \texttt{NOAA\_AR} is the NOAA active-region number; and \texttt{From\_SMARP} indicates whether a record contains SMARP-derived values. The magnetic-field and active-region variables are defined in Table~\ref{tab:sharp_parameters}.

In \texttt{Flare.csv}, \texttt{Duration} and \texttt{RisenTime} are measured in minutes; \texttt{Strength} is flare peak flux in W\,m$^{-2}$; and \texttt{log\_Strength} is the base-10 logarithm of \texttt{Strength}. In \texttt{CDAWDONKI\_CME.csv}, \texttt{cme\_time\_posix} denotes CME event time; \texttt{lon\_deg}, \texttt{halfAngle\_deg}, and \texttt{speed\_km\_s} denote longitude, half angular width, and speed, respectively; and \texttt{From\_CDAW} indicates whether CDAW information contributed to the record. In \texttt{CDAW\_CME.csv}, \texttt{central\_pa\_deg}, \texttt{width\_deg}, \texttt{speed\_km\_s}, and \texttt{energy\_erg} denote central position angle, angular width, speed, and kinetic energy, respectively.

In \texttt{GOESHAPI\_ProtonFlux.csv}, \texttt{time\_posix} is the measurement time, \texttt{ProtonFlux} is the $>$10~MeV proton flux in pfu, and \texttt{goes} identifies the contributing GOES satellite. In \texttt{Merged\_XRays.csv}, \texttt{time\_posix} is the measurement time, \texttt{xrsb} is the 0.1--0.8~nm X-ray flux in W\,m$^{-2}$, and \texttt{goes} identifies the contributing GOES satellite.

\paragraph{Analyzed Data.}
The \texttt{Analyzed Data/} directory contains two model-ready feature tables listed in Table \ref{tab:analyzed-inventory}. 

\begin{table}[htbp]
  \centering
  \caption{Files in \texttt{Analyzed Data/}.}
  \label{tab:analyzed-inventory}
  \begin{tabular}{@{}p{7cm}p{1cm}p{6cm}@{}}
    \toprule
    File & Step  & Use case \\
    \midrule
    \texttt{rolling\_combinded\_seq\_24hours.csv}
      & 24 h  & Primary non-overlapping daily table \\
    \rowcolor{lightgray!30}\texttt{rolling\_combinded\_seq\_1hours.csv}
      & 1 h & High-resolution overlapping windows \\
    \bottomrule
  \end{tabular}
\end{table}

In both files, \texttt{window\_begin} and \texttt{window\_end} identify the predictor-window boundaries. Variables prefixed with \texttt{Future\_} are targets associated with the future prediction interval. Variables ending in \texttt{\_label} are binary indicators; variables ending in \texttt{\_num} are event counts; and variables following the convention \texttt{\{Source\}\_\{Feature\}\_\{min|max|avg\}} are window-level minimum, maximum, or mean summaries. The prefixes \texttt{SHARP\_} and \texttt{SHARP\_AR\_} distinguish summaries of all active-region records from summaries of flare-associated active-region records, respectively. Table \ref{tab:future-targets} defines the future-window target variables.



\begin{table}[htbp]
  \centering
  \caption{Future-window target variables in the analyzed data files.}
  \label{tab:future-targets}
  \small
  \begin{tabular}{@{}llp{8cm}@{}}
    \toprule
    Variable & Type & Definition \\
    \midrule
    \texttt{Future\_OSEP\_label} & Binary & 1 if an operational SEP ($\geq 10$ pfu) occurs in the next 24 h \\
    \rowcolor{lightgray!30}\texttt{Future\_GSEP\_label} & Binary & 1 if a general SEP ($\geq 10^{-6}$ pfu) occurs in the next 24 h \\
    \texttt{Future\_Flare\_num} & Count & Number of flares in the next 24 h \\
    \rowcolor{lightgray!30}\texttt{Future\_DONKICME\_num} & Count & Number of fused DONKI CMEs in the next 24 h \\
    \texttt{Future\_CDAWCME\_num} & Count & Number of CDAW CMEs in the next 24 h \\
    \rowcolor{lightgray!30}\texttt{Future\_DONKICME\_speed\_km\_s\_max} & Continuous & Maximum fused CME speed in the next 24 h \\
    \texttt{Future\_log\_DONKICME\_speed\_km\_s\_max} & Continuous & $\log_{10}(\text{peak speed} + 10^{-4})$ \\
    \rowcolor{lightgray!30}\texttt{Future\_ProtonFlux\_max} & Continuous & Peak GOES $>$10 MeV flux (pfu) in the next 24 h \\
    \texttt{Future\_log\_ProtonFlux\_max} & Continuous & $\log_{10}$ of peak proton flux \\
   \rowcolor{lightgray!30} \texttt{Future\_XRSB\_max} & Continuous & Peak 0.1-0.8 nm X-ray flux (W\,m$^{-2}$) \\
    \texttt{Future\_log\_XRSB\_max} & Continuous & $\log_{10}$ of peak X-ray flux \\
    \bottomrule
  \end{tabular}
\end{table}

\paragraph{Code.}
The \texttt{Code/} directory includes \texttt{fetch\_solar\_data.ipynb} for source-data retrieval; the \texttt{Dataset\_Preprocessing/} subdirectory for source-specific preprocessing and data fusion; \texttt{All\_features\_RData.R} and \texttt{Data\_Aggregation.R} for generating the processed feature workspace and analyzed data files; and the \texttt{Plot/} subdirectory for publication-figure generation.

\section*{Data Overview}

{SEP-PRISM Data spans 3 February 1986 through 10 September 2025 and integrates SEP events, solar flares, active-region magnetic parameters, CME catalogs, GOES soft X-ray flux, and GOES $>$10~MeV proton flux. The released processed and analyzed data products cover event-based catalogs and continuous time series sampled at their native or harmonized cadences. Flare, CME, and SEP records are event-based; the merged GOES X-ray and proton-flux series have a 5-minute cadence; SMHARP records have a 96-minute cadence before the SHARP era and a 12-minute cadence for native SHARP observations after May 2010.}

{The local source-data inputs used to construct the released SEP-PRISM Data products comprise 569 CLEAR SEP Benchmark entries, 78,724 flare records, 4,889,762 SHARP records, 68,166 SMARP records, 8,366 DONKI CME records, and 16,216 CDAW CME records. The unified SMHARP archive contains 5,310,881 records, and the unified CDAWDONKI CME archive contains 17,463 records.}

{The analyzed data products each contain 274 columns. The non-overlapping daily file, \texttt{rolling\_combinded\_seq} \texttt{\_24hours.csv}, contains 14,464 24-hour predictor windows. Of these windows, 650 (4.5\%) have \texttt{Future\_OSEP\_label}=1, and 2,122 (14.7\%) have \texttt{Future\_GSEP\_label}=1. The higher-frequency file, \texttt{rolling\_combinded\_seq\_1hours.csv}, contains 347,136 24-hour windows advanced at 1-hour increments.}

{The dataset includes both native and reconstructed records. The \texttt{From\_SMARP} and \texttt{From\_CDAW} provenance flags identify records in which aligned SMARP-derived or CDAW-derived information contributes to the corresponding unified archive. These flags enable users to restrict analyses to native records or to evaluate sensitivity to the inclusion of reconstructed values.}

\section*{Technical Validation}

Technical quality and reuse potential were assessed through four complementary checks: evaluation of the SHARP-SMARP fusion during their 2010 overlap period, evaluation of the DONKI-CDAW CME fusion during 2010-2025, verification that temporal coverage was extended using aligned surrogate records rather than by truncating the historical archive, and confirmation
that all predictor families were mapped into a uniform rolling-window schema. 

\subsection*{SHARP-SMARP fusion validation}
The SHARP-SMARP alignment was evaluated over the 2010 overlap period by comparing native SHARP observations with SMARP-derived estimates using Pearson correlation coefficients (COR) and root mean square error (RMSE). {COR measures linear association, whereas RMSE measures absolute differences in scale.} Results are summarised in Table \ref{tab:alignment_performance}.

For parameters common to both catalogs, raw SHARP-SMARP correlations are generally high (typically COR $>$ 0.95), consistent with shared physical definitions and correlated temporal variability. Nevertheless, several shared fields, including \texttt{USFLUXL}, \texttt{MEANGBL}, \texttt{CMASKL}, \texttt{RSUN\_OBS}, and \texttt{DSUN\_OBS}, show appreciable magnitude offsets. These discrepancies likely reflect instrumental and processing differences between SOHO/MDI and SDO/HMI, which can produce scale inconsistencies even when temporal co-variation remains strong.

\begin{figure}[hpbt]
    \centering
    \includegraphics[width=1\linewidth]{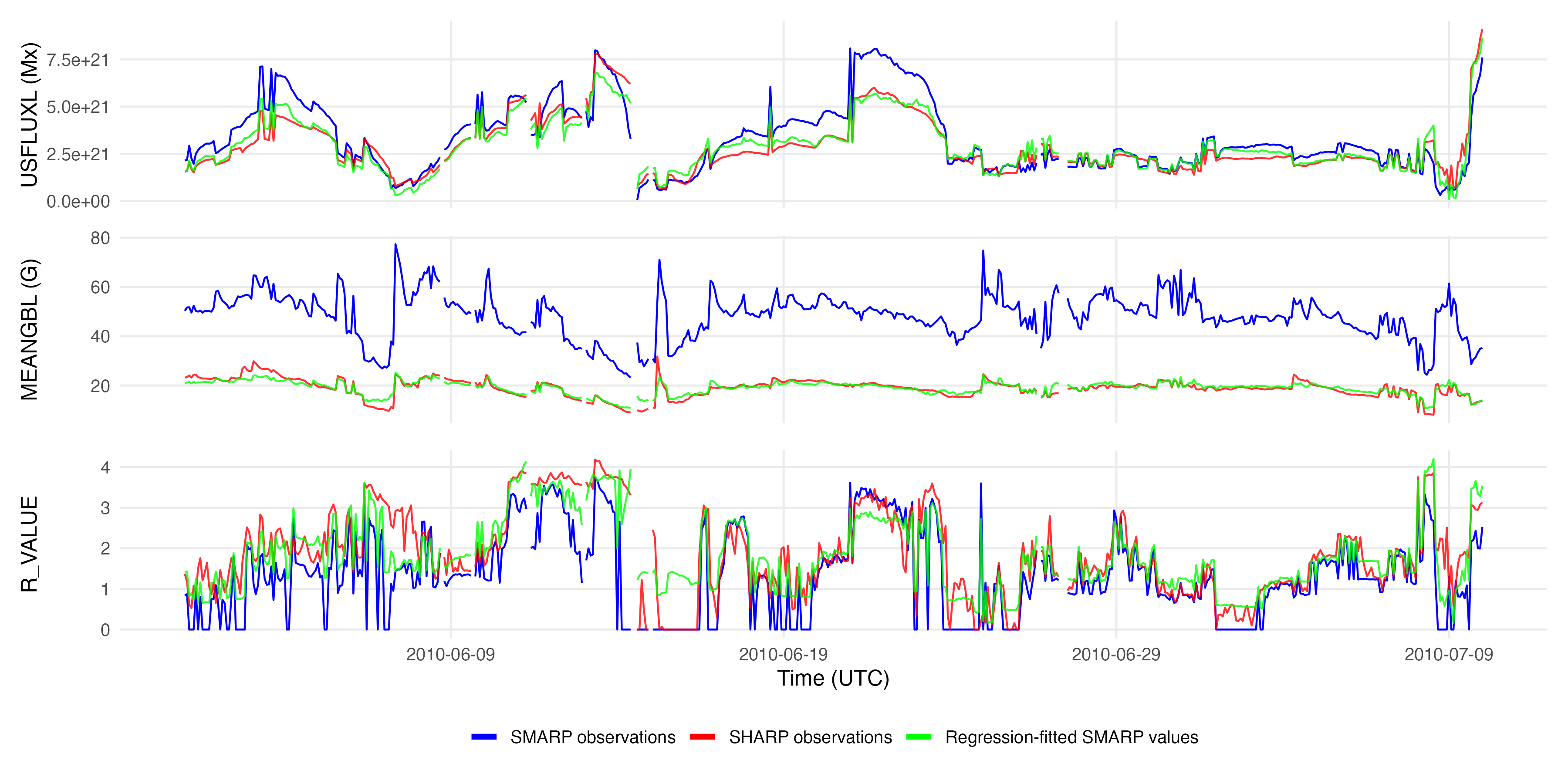}
    \caption{Comparison of SMARP observations, SHARP observations, and regression-based fitted SHARP values for three shared parameters over the period from 1 June 2010 to 10 July 2010.  }
    \label{fig:SHARP_SMARP}
\end{figure}

After regression-based alignment, both COR and RMSE improve for most shared parameters. For example, \texttt{USFLUXL} increases from COR $=$ 0.9605 to 0.9849 with a corresponding RMSE reduction; \texttt{MEANGBL} improves from COR $=$ 0.7689 to 0.8433, with RMSE falling from 30.19 to 2.55; and \texttt{R\_VALUE} rises from COR $=$ 0.8336 to 0.9016, with RMSE decreasing from 1.06 to 0.72. Figure \ref{fig:SHARP_SMARP} shows the corresponding aligned time series for these three parameters over the period from 1 June 2010 to 10 July 2010. Similar gains are observed for size and morphology descriptors (\texttt{NPIX}, \texttt{SIZE}, \texttt{AREA}). Positional quantities (\texttt{LAT\_MIN}, \texttt{LON\_MIN}, \texttt{LAT\_MAX}, \texttt{LON\_MAX}, \texttt{LAT\_FWT}, \texttt{CRLT\_OBS}) already show near-unity agreement before alignment (COR $\approx$ 1); alignment mainly suppresses small residual errors, as indicated by further RMSE reductions.

For SHARP-only vector-field parameters without direct SMARP counterparts, only post-alignment performance is reported. Integrated quantities such as \texttt{USFLUX} (COR $=$ 0.9689), \texttt{TOTUSJZ} (COR $=$ 0.9646), and \texttt{TOTPOT} (COR $=$ 0.9455) are reconstructed with high fidelity. Moderate agreement is obtained for shear- and energy-related fields, including \texttt{MEANPOT} (COR $=$ 0.7459), \texttt{MEANSHR} (COR $=$ 0.7597), and \texttt{SHRGT45} (COR $=$ 0.7737). Lower correlations (COR $\approx$ 0.44-0.50) are found for derivative and helicity-related measures such as \texttt{MEANJZD}, \texttt{MEANALP}, and \texttt{MEANJZH}, consistent with the greater information content required to infer vector magnetic properties from line-of-sight data. {Because native SHARP measurements are unavailable before 2010, reconstructed \texttt{MEANJZD}, \texttt{MEANALP}, and \texttt{MEANJZH} values are more uncertain than other SMHARP parameters. Analyses using these descriptors for pre-2010 samples should therefore be interpreted cautiously; the \texttt{From\_SMARP} provenance flag can be used to isolate reconstructed records in sensitivity analyses. Agreement was higher for most other shared and SHARP-specific parameters.}

\begin{longtable}{lll}
\caption{Pearson correlation coefficients and root mean square errors (RMSE; values in parentheses) between SHARP observations and SMARP-derived estimates during the 2010 overlap period, before and after alignment. Parameters marked with $^\ast$ are shared between SHARP and SMARP. For SHARP-specific parameters, only post-alignment results are reported.}
\label{tab:alignment_performance} \\
\hline
Feature & Before & After \\ \hline
\texttt{USFLUX}  &\qquad\qquad -- & 0.9689 (1.7367+21) \\
\rowcolor{lightgray!30}\texttt{MEANGAM} & \qquad\qquad -- & 0.7414 (7.2543) \\
\texttt{MEANGBT} &\qquad\qquad -- & 0.8185 (1.6299+1) \\
\rowcolor{lightgray!30}\texttt{MEANGBZ} &\qquad\qquad -- & 0.7944 (1.6390e+1)\\
\texttt{MEANGBH} &\qquad\qquad -- & 0.7387 (9.3501)\\
\rowcolor{lightgray!30}\texttt{MEANJZD} &\qquad\qquad -- & 0.4572 (0.4153)\\
\texttt{TOTUSJZ} &\qquad\qquad -- & 0.9646 (2.0781e+12)\\
\rowcolor{lightgray!30}\texttt{MEANALP} & \qquad\qquad -- & 0.4484 (0.0173) \\
\texttt{MEANJZH} &\qquad\qquad  -- & 0.5088 (0.0056)\\
\rowcolor{lightgray!30}\texttt{SAVNCPP} &\qquad\qquad -- & 0.7717 (1.4815e+12)\\
\texttt{MEANPOT} &\qquad\qquad -- & 0.7459 (3.3833e+3) \\
\rowcolor{lightgray!30}\texttt{TOTPOT} &\qquad\qquad -- & 0.9455 (3.5697e+22)\\
\texttt{MEANSHR} &\qquad\qquad -- & 0.7597 (6.3913)\\
\rowcolor{lightgray!30}\texttt{SHRGT45} &\qquad\qquad -- & 0.7737 (1.2133e+1)\\
\texttt{USFLUXL$^\ast$} & 0.9605 (2.0521e+21) & 0.9849 (8.7157e+20)\\
\rowcolor{lightgray!30}\texttt{MEANGBL$^\ast$} & 0.7689 (3.0186e+1)& 0.8433 (2.5516) \\
\texttt{NPIX$^\ast$}    & 0.9608 (5.6356e+4) & 0.9725 (9.9283e+3) \\
\rowcolor{lightgray!30}\texttt{SIZE$^\ast$}    & 0.9607 (1.4620e+3)& 0.9723 (8.9127e+2) \\
\texttt{AREA$^\ast$}    & 0.9496 (9.9709e+2) & 0.9652 (6.0350e+2)\\
\rowcolor{lightgray!30}\texttt{NACR$^\ast$}     & 0.9812 (5.6493e+3) & 0.9895  (6.6011e+2)\\
\texttt{SIZE\_ACR$^\ast$} & 0.9812 (1.2444e+2) & 0.9894 (5.9265e+1) \\
\rowcolor{lightgray!30}\texttt{AREA\_ACR$^\ast$} & 0.9690 (9.5493e+1) & 0.9849 (4.4182+1)\\
\texttt{LAT\_MIN$^\ast$}  & 0.9982 (1.4158)& 0.9985 (1.1751)\\
\rowcolor{lightgray!30}\texttt{LON\_MIN$^\ast$}  & 0.9983 (2.4752) & 0.9985 (2.0929) \\
\texttt{LAT\_MAX$^\ast$}  & 0.9984 (1.4050) & 0.9986 (1.1262) \\
\rowcolor{lightgray!30}\texttt{LON\_MAX$^\ast$}  & 0.9981 (2.7312) & 0.9987 (1.9876)\\
\texttt{LAT\_FWT$^\ast$}  & 0.9994 (0.7299) & 0.9995 (0.6743)    \\
\rowcolor{lightgray!30}\texttt{LON\_FWT$^\ast$}  & 0.9995 (1.2802) & 0.9995 (1.1702)  \\
\texttt{CRLT\_OBS$^\ast$} & 1.0000 (0.0368) & 1.0000 (0.0096)  \\
\rowcolor{lightgray!30}\texttt{CRLN\_OBS$^\ast$} & 0.9927 (1.2048+1) & 0.9927 (1.2003e+1)    \\
\texttt{R\_VALUE$^\ast$}  & 0.8336 (1.0583) & 0.9016 (0.7181) \\
\rowcolor{lightgray!30}\texttt{CMASKL$^\ast$}    & 0.9485 (7.4431e+4) & 0.9654 (1.3672e+4)  \\
\texttt{DSUN\_OBS$^\ast$} & 0.9925 (1.4565e+9) & 0.9996 (2.6590e+7)  \\
\rowcolor{lightgray!30}\texttt{RSUN\_OBS$^\ast$} & 0.9927 (9.2433) & 0.9996 (0.1676)  \\
\hline
\end{longtable}

\subsection*{DONKI-CDAW CME fusion validation}
The DONKI-CDAW alignment was assessed over 3 April 2010 to 9 September 2025 using COR and RMSE computed from temporally matched event pairs (Table \ref{tab:CME_alignment_performance}). For the shared variable \texttt{speed\_km\_s}, the raw CDAW-DONKI correlation is moderate (COR $=$ 0.5562), plausibly reflecting differences in measurement methodology and projection treatment between catalogs. After regression-based reconstruction, COR increases to 0.6479 and RMSE decreases, indicating improved though still imperfect speed agreement. For DONKI-specific descriptors, only post-alignment performance is evaluated. Longitude (\texttt{lon\_deg}; COR $=$ 0.4899) and half-angle (\texttt{halfAngle\_deg}; COR $=$ 0.4560) show moderate reconstruction skill, whereas latitude (\texttt{lat\_deg}; COR $=$ 0.0821) is poorly recovered and is associated with large RMSE. The weak latitude performance likely reflects observational-geometry effects and catalog-specific definitions; consequently, \texttt{lat\_deg} was omitted from the final CME feature set used in downstream analysis. {These results indicate that the alignment procedure achieves moderate-to-good reconstruction of the shared speed variable and of the DONKI-specific longitude and half-angle descriptors; latitude was excluded because of its low cross-catalog agreement.}

\begin{longtable}{lll}
\caption{Pearson correlation coefficients and RMSE (values in parentheses) between DONKI CME observations and CDAW-derived estimates during 3 April 2010 to 9 September 2025, before and after alignment. Features marked with $^\ast$ are shared between catalogs. For DONKI-specific features, only post-alignment results are reported.}
\label{tab:CME_alignment_performance} \\
\hline
Feature & Before & After \\ \hline
\texttt{lat\_deg}  &\qquad\qquad -- & 0.0821 (3.1002e+1) \\
\rowcolor{lightgray!30}\texttt{lon\_deg}  & \qquad\qquad -- & 0.4899 (8.5917e+1) \\
\texttt{halfAngle\_deg}  &\qquad\qquad -- & 0.4560 (1.0782e+1) \\
\rowcolor{lightgray!30}\texttt{speed\_km\_s$^\ast$} & 0.5562 (3.0106e+2) & 0.6479 (2.4769e+2)  \\
\hline
\end{longtable}

\subsection*{Temporal coverage and windowing consistency}
Temporal continuity was preserved by extending early-era records using aligned surrogate features rather than truncating the dataset to the years with full native SHARP or DONKI availability. The provenance flags \texttt{From\_SMARP} and \texttt{From\_CDAW} indicate when reconstructed values replace direct observations, allowing users to filter or stratify analyses by data origin.

All predictor families were aggregated {using a consistent 24-hour rolling-window schema, with each record summarizing observations from the preceding 24 hours. The analyzed data include both non-overlapping 24-hour windows, advanced at 24-hour intervals, and overlapping 24-hour windows, advanced at 1-hour intervals; see table \ref{tab:analyzed-inventory}. Within each window, variables were summarized using harmonized minimum, maximum, and mean statistics and assigned consistent column names.} This design reduces cross-source format heterogeneity and yields directly comparable samples across solar cycles and feature domains. Independent modeling experiments reported in the companion article provide an additional, non-primary indication that the curated records form a coherent predictive feature set \cite{yu2026OperationalSEP}. In those analyses, combinations of magnetic, proton-flux, and radiative inputs yielded the strongest performance in classification and auxiliary regression.

\section*{Usage Notes}

The SEP Prediction Database, \texttt{SEP-PRISM Data}, is intended to support research on SEP forecasting, solar-driver analysis, and machine-learning benchmark development. The archive provides three workflow tiers for different reuse scenarios. {The \texttt{Raw Data/} directory is retained as a local working directory for users who wish to retrieve, inspect, or modify source data acquisition and fusion procedures. Downloaded external raw records are not included in the public Zenodo archive; instead,
they are obtained directly from the original providers when the retrieval workflow is executed.} The \texttt{Processed Data/} directory contains harmonized source-derived tables for event-level and time-series analyses at native or harmonized cadence, including 5-minute flux series, pre-SHARP-era SMHARP records, 12-minute native SHARP-era records, and event-based flare and CME records. The \texttt{Analyzed Data/} directory contains model-ready data products for supervised learning and operational forecasting experiments using pre-aggregated 24-hour predictor windows and 24-hour future prediction windows.

Most users seeking a model-ready table should begin with \texttt{rolling\_combinded\_seq\_24hours.csv} which contains 14,464 rows and 274 columns. The companion file \texttt{rolling\_combinded\_seq\_1hours.csv}, provides overlapping 24-hour predictor windows advanced in 1-hour increments. This file supports higher-frequency forecasting analyses but is substantially larger (approximately 802 MB) and contains correlated samples; users should therefore use cross-validation schemes that account for temporal dependence and
overlapping windows.

For direct use, \texttt{Analyzed Data/rolling\_combinded\_seq\_24hours.csv} can be loaded in any CSV-compatible environment such as R, Python, or MATLAB. Each row defines a past 24-hour feature window and a future 24-hour prediction horizon. The primary classification target is \texttt{Future\_OSEP\_label}. The broader auxiliary target \texttt{Future\_GSEP\_label} is retained to support methodological development and multi-task learning experiments. 

For full reproducibility, the pipeline can be rebuilt from raw inputs by downloading or cloning the repository, placing the CLEAR Benchmark SEP catalog in \texttt{Raw Data/}, setting \texttt{REPO\_ROOT} in the scripts, and running the preprocessing scripts in followed by \texttt{All\_features\_RData.R} and \texttt{Data\_Aggregation.R}. The notebook \texttt{Code/fetch\_solar\_data.ipynb} can be used to refresh raw downloads and near-real-time data fetching. To generate 1-hour stepped windows, set \texttt{FORWARD\_HOURS <- 1L} in \texttt{Data\_Aggregation.R} before execution.

Reprocessing requires R version 4.5.1 with packages including \textit{data.table}, \textit{lubridate}, \textit{readr}, \textit{MASS}, \textit{VIM}, and \textit{dplyr}. The acquisition notebook requires Python version 3.12.3 with \textit{sunpy}, \textit{drms}, and related dependencies. Exact version information should be preserved in the release notes associated with the archived dataset version.

Several limitations should be considered before reuse. First, pre-2010 magnetic predictors and pre-2010 DONKI-like CME descriptors are statistically reconstructed surrogate variables rather than native observations. Second, SEP labels and source associations depend on external catalogs and may change if those catalogs are revised. Third, the analyzed tables compress timing and spatial structure into window-level minimum, maximum, and mean summaries. Applications requiring event-level timing, spatial mapping, or sub-hourly lead times should use the processed data products or reconstruct the relevant records from the original source providers. Fourth, SEP-PRISM Data is a curated research archive ending on 10 September 2025; it is not itself a continuously updated real-time operational product.

\section*{Data Availability}

The curated, processed, and analyzed SEP-PRISM Data products generated in this study are publicly available from Zenodo \cite{Yu2026SEPPRISMData} at \url{https://doi.org/10.5281/zenodo.21297634}. Downloaded third-party raw input data are not redistributed in the Zenodo archive. The CLEAR SEP Benchmark catalog is available from the public \texttt{fetchsep} GitHub release, and the remaining third-party source data are available from their original public providers. The Zenodo archive includes the retrieval notebook, preprocessing workflow, source endpoints, product identifiers, query specifications, and documentation required to obtain the source records and reproduce the released SEP-PRISM Data products.

\section*{Code Availability}

All custom code developed for source-data retrieval, preprocessing, source alignment, feature construction, technical validation, and figure generation is publicly available from the Zenodo archive associated with this dataset \cite{Yu2026SEPPRISMData} at \url{https://doi.org/10.5281/zenodo.21297634}. The archive contains
\texttt{Code/fetch\_solar\_data.ipynb}; source-specific R scripts in \texttt{Code/Dataset\_Preprocessing/};
\texttt{Code/All\_features\_RData.R}; \texttt{Code/Data\_Aggregation.R}; and the plotting scripts in
\texttt{Code/Plot/}. The retrieval notebook obtains third-party source records directly from their original providers and writes them to the local \texttt{Raw Data/} working directory; downloaded raw files are not included in the public archive.

\section*{{Funding}}
This work is supported by the NASA Space Weather Center of Excellence program under award numbers 80NSSC23M0191 and 80NSSC23M0192.

\section*{Author Contributions Statement}
Y.C. conceived the study and revised the manuscript. Y.Y. performed the data collection and preparation, conducted the technical validation, and drafted the manuscript. K.W. performed the SEP events collection and participated in manuscript revision. L.Z., W.M., and T.G. contributed to the scientific interpretation and revision of the manuscript. All authors reviewed the manuscript.

\section*{Competing Interests}
The authors declare no competing interests.

\bibliography{sample}
\end{document}